\documentclass[pdflatex,sn-mathphys-num]{sn-jnl}% Math and Physical Sciences Numbered Reference Style
\usepackage{graphicx}%
\usepackage{multirow}%
\usepackage{amsmath,amssymb,amsfonts}%
\usepackage{amsthm}%
\usepackage{mathrsfs}%
\usepackage[title]{appendix}%
\usepackage{xcolor}%
\usepackage{textcomp}%
\usepackage{manyfoot}%
\usepackage{booktabs}%
\usepackage{algorithm}%
\usepackage{algorithmicx}%
\usepackage{algpseudocode}%
\usepackage{listings}%
\theoremstyle{thmstyleone}%
\theoremstyle{thmstyletwo}%

\theoremstyle{thmstylethree}%

\begin{document}

\title[Article Title]{Complexity of Financial Time Series:  Multifractal and Multiscale Entropy Analyses}

%%=============================================================%%
%% GivenName	-> \fnm{Joergen W.}
%% Particle	-> \spfx{van der} -> surname prefix
%% FamilyName	-> \sur{Ploeg}
%% Suffix	-> \sfx{IV}
%% \author*[1,2]{\fnm{Joergen W.} \spfx{van der} \sur{Ploeg} 
%%  \sfx{IV}}\email{iauthor@gmail.com}
%%=============================================================%%

\author*[1]{\fnm{Oday} \sur{Masoudi}}\email{adimasoudi1998@gmail.com}

\author[1]{\fnm{Farhad} \sur{Shahbazi}}\email{shahbazi@iut.ac.ir}

\author[1]{\fnm{Mohammad} \sur{Sharifi}}\email{mohammad.sharifi.baraftab@gmail.com}

\affil*[1]{\orgdiv{Department of Physics}, \orgname{Isfahan University of Technology},  \state{Isfahan} \postcode{84156-83111}, \country{Iran}}

%\affil[2]{\orgdiv{Department}, \orgname{Organization}, \orgaddress{\street{Street}, \city{City}, \postcode{10587}, \state{State}, \country{Country}}}

%\affil[3]{\orgdiv{Department}, \orgname{Organization}, \orgaddress{\street{Street}, \city{City}, \postcode{610101}, \state{State}, \country{Country}}}

%%==================================%%
%% Sample for unstructured abstract %%
%%==================================%%

\abstract{We employed Multifractal Detrended Fluctuation Analysis (MF-DFA) and Refined Composite Multiscale Sample Entropy (RCMSE) to investigate the complexity of Bitcoin, GBP/USD, gold, and natural gas price log-return time series. This study provides a comparative analysis of these markets and offers insights into their predictability and associated risks. Each tool presents a unique method to quantify time series complexity. The RCMSE and MF-DFA methods demonstrate a higher complexity for the Bitcoin time series than others. It is discussed that the increased complexity of Bitcoin may be attributable to the presence of higher nonlinear correlations within its log-return time series.}

\keywords{Multi-scale Entropy, Multi-fractal, Sample Entropy, Time series Analysis}
%%\pacs[JEL Classification]{D8, H51}

%%\pacs[MSC Classification]{35A01, 65L10, 65L12, 65L20, 65L70}

\maketitle

\section{Introduction}\label{sec1}
The measurement of time-series complexity has prompted numerous efforts in the scientific community, thereby motivating the development of tools for quantifying complexity. Various methodologies have been established for this purpose, including the Lyapunov exponent, the fractal dimension, and entropy \cite{bib1}. Entropy, as a metric of complexity, is a popular approach that quantifies the average amount of information or uncertainty present within a system \cite{bib2}. Shannon first introduced the application of entropy to signals and time series in 1948 in his seminal work \textit{"A Mathematical Theory of Communication"} \cite{bib3}. 

Several approaches based on Shannon's theory have been proposed to measure the entropy of time series. These include approximate entropy, sample entropy \cite{bib4}, and permutation entropy \cite{bib5}. 
Many researchers have applied entropy definitions to study a system's behavior. For instance, in one study, the authors used wavelet entropy and complexity-entropy curves to examine the predictability of energy commodity prices \cite{bib6}. 
However, in the traditional method of measuring the regularity of a series, higher complexity was reported for a system with a higher entropy. Nevertheless, an increase in entropy does not necessarily indicate an increase in dynamical complexity. 

Costa introduced a novel method for quantifying the complexity of time series in 2002. In this method of measuring complexity, multiple time scales are considered instead of a single one, which is achieved through the coarse-graining procedure of the original time series \cite{bib7}.  
The complexity analysis of time series across multiple scales has been extensively explored in various research areas using entropy. For example, Jing Wang utilized multiscale entropy analysis (MSE) and its modified version, multiscale permutation entropy analysis (MSPE), to quantify the complexity of traffic data in weekend and workday time series, aiding in classification and prediction \cite{bib8}. Costa et al. used the MSE method to measure biological signal complexity and distinguish between various disease states. Furthermore, the method was employed to analyze DNA sequences, revealing higher multiscale entropy in non-coding DNA, indicating biological significance beyond traditional understanding \cite{bib9}. MSE has also been widely used in biology and psychological research \cite{bib10, bib11, bib12}.

Jianan and Pengjian utilized multiscale sample entropy to analyze markets across Asia, Europe, and America. They found that European markets show lower entropy than Asian markets, but higher entropy than those in the American markets \cite{bib13}. In another research, the authors used a multiscale entropy-based approach to examine the interactions between innovative financial assets (e.g., cryptocurrencies, green bonds) and traditional assets (e.g., gold, crude oil, and government bonds) across multiple time scales. The results revealed heterogeneity in information exchange intensity, with innovative assets dominating in the short term and traditional assets dominating in the long term \cite{bib14}. The researchers used multiscale approximate entropy in a separate study to explore market efficiency in foreign exchange (FX) markets. They found that developed FX markets are more efficient than emerging FX markets \cite{bib15}.
Several studies have also been done on the behavior of financial market time series using approximate and sample entropy \cite{bib16, bib17, bib18, bib19}. 

Another method of measuring time series complexity is the detrended fluctuation analysis (DFA) method, introduced by Peng in 1993 \cite{bib20}. This method determines fractal scaling properties and detects long-range correlations in noisy, nonstationary time series. However, in many cases, using a simple monofractal scaling behavior is insufficient to describe the different aspects of the series' behavior. Therefore, it is necessary to apply multiple scaling exponents to provide a more comprehensive description of the scaling behavior. The concept of utilizing multifractal analysis was first introduced by Kantelhardt et al. \cite{bib21}. They developed Multifractal Detrended Fluctuation Analysis (MF-DFA), an extension of DFA to determine fractal exponents at different scales in non-stationary time series. The method differentiates between multifractality arising from long-range correlations and that resulting from a broad probability density function.

The intricate relationships among economic policy uncertainty (EPU), the crude oil market, and the stock market have been thoroughly examined employing multifractal methods. It has been determined that all three markets exhibit multifractality, which is driven by long-range correlations and fat-tailed distributions. The study elucidates robust cross-correlations, particularly between stock and oil prices, and demonstrates that the stock market exerts the most significant influence on these relationships, whereas EPU has the least impact \cite{bib22}. 
In another study, the complexity analysis was conducted on high-frequency intraday data derived from six major cryptocurrencies (Bitcoin, Ethereum, Litecoin, Dashcoin, EOS, and Ripple) and six prominent forex markets (Euro, British pound, Canadian dollar, Australian dollar, Swiss franc, and Japanese yen) utilizing the MFDFA method \cite{bib23}. This analysis yielded significant multifractal properties across all series, exhibiting varying multifractal strengths, particularly emphasizing Dashcoin's higher multifractal strength and Litecoin's lower multifractal strength.

This research aims to apply Multifractal Detrended Fluctuation Analysis (MF-DFA) and Refined Composite Multiscale Sample Entropy (RCMSE) to overcome the limitations of traditional multiscale sample entropy methods in analyzing financial time series. Firstly, the time series must be sufficiently lengthy to be applicable in the MSE method. Unfortunately, many financial time series are short, which complicates the acquisition of adequately lengthy data for this purpose. Consequently, the use of MSE may yield inaccurate results. Secondly, within the MSE methodology, as the scale factor increases, the probability of encountering undefined entropy for the coarse-grained time series also rises \cite{bib24}. Therefore, we have chosen to utilize the RCMSE algorithm, which is more suitable for short time series. This method, introduced by Shuen-De Wu in 2014, was specifically designed to extend multiscale entropy analysis to short time series \cite{bib25}.
We believe that the combination of MF-DFA and RCMSE provides valuable insight into both the predictability and fractal characteristics of financial markets, which are two fundamental features of these systems.

The organization of this paper is as follows: Section 2 discusses the methodology for computing multiscale entropy analysis, its refined version, and multifractal detrended fluctuation analysis. Section 3 describes the data utilized in this research and their respective sources. Section 4 presents the results of both the RCMSE and MF-DFA analyses conducted on the provided empirical data, while Section 5 concludes the research.

\section{Methods}\label{sec2}
\subsection{Sample Entropy}
\label{subsec1}
Sample entropy is a method used to evaluate the likelihood that patterns of data points, which are similar over a defined length, remain similar when one additional point is added. Let's discuss the procedure for computing the sample entropy of a time series \cite{bib4}. Assuming the time series is represented by $x=\{x_1, \dots, x_N\}$ with $N$ points, sample entropy of the series is computed according to the following steps:

Step 1) Create $N-m+1$ overlapping embedding vectors $X_m(i)$ with the length of m with the following condition:
\begin{equation}
X_m(i) = \{x_i, x_{i+1}, \dots, x_{i+m-1}\}, \quad 1 \leq i \leq N - m+1.
\label{eq:embedding_vector}
\end{equation}
Which in this equation m is embedding dimension. 

Step 2) Determining the distance between two vectors using the equation \eqref{eq:distance}:
\begin{equation}
    d[X_m(i), X_m(j)] = \max \{ |x_{i+k} - x_{j+k}| : 0 \leq k \leq m-1 \}
    \label{eq:distance}
\end{equation}
In this equation, $\quad 1 \leq i, j \leq N-m+1, \, i \neq j$ and the distance is defined as the maximum difference of their corresponding scalar components.

Step 3) Let $n^m$ represent the total number of $m$-dimensional matched vectors $d[X_m(i), X_m(j)]$ that are less than or equal to a tolerance $r$. Which $r$ is commonly considered a fraction of the standard deviation of the time series.

Step 4) For $m = m+1$, repeat steps 1-3, and let $n^{m+1}$ represent the number of $m+1$ dimensional matched vectors.

Step 5) The sample entropy is then defined by the equation \eqref{eq:SampEn}:
\begin{equation}
    \text{SampEn}(x, m, r) = -\ln\left(\frac{n^{m+1}}{n^m}\right)
    \label{eq:SampEn}
\end{equation}

The SampEn algorithm might encounter issues with short time series. If either $n^{(m+1)}$ or $n^m$ is zero, it results in undefined entropy. To ensure a meaningful sample entropy, the length of the time series should fall within the range of $10^m$ to $30^m$ \cite{bib26}.

\subsection{Multiscale Sample Entropy (MSE)}
\label{subsec2}
\begin{figure}[H] 
\centering
\includegraphics[width=0.7\textwidth]{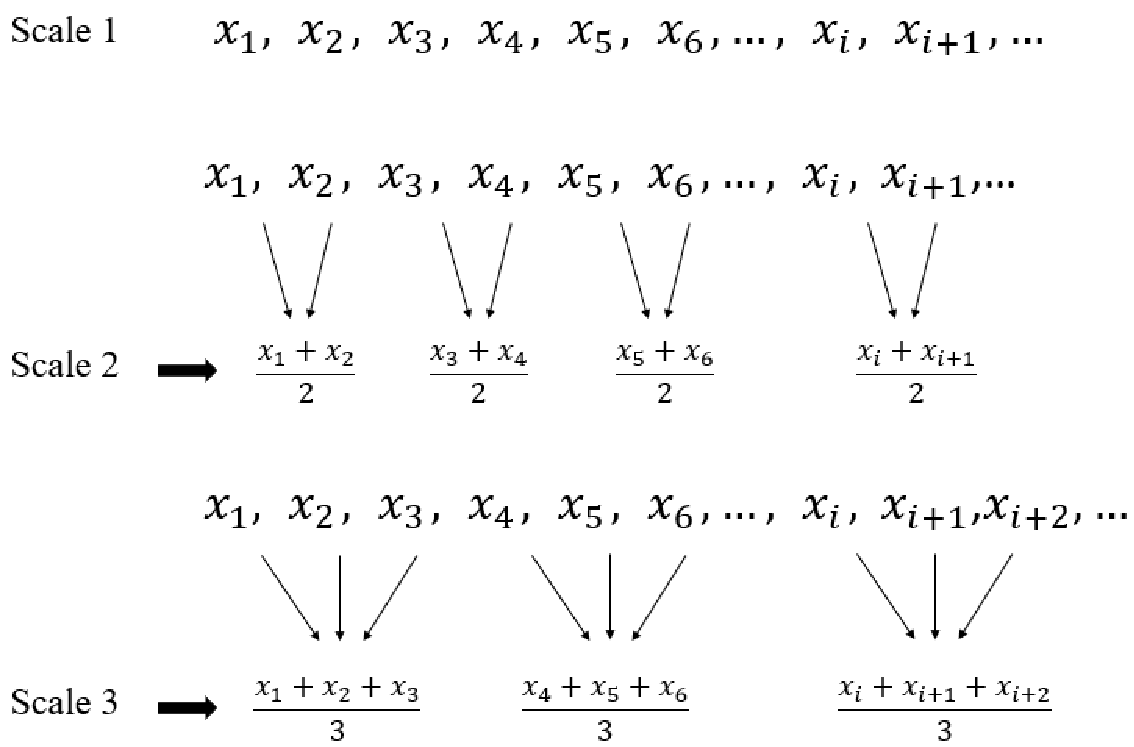} % Replace 
\caption{Illustration of the coarse graining procedure.}
\label{fig:multiscaling}
\end{figure}

Using multiple scales instead of a single scale in entropy analysis first was introduced by Costa in 2002 \cite{bib7}. These multiscales are achieved by a coarse-graining procedure, and the number of scales is determined by scale factor of $\tau$. For a given time series $x=\{x_1, \dots, x_N\}$ coarse-grained series are computed using Equation \eqref{eq:coarse_grain}.
\begin{equation}
y_j^{(\tau)} = \frac{1}{\tau} \sum_{i=(j-1)\tau+1}^{j\tau} x_i, \quad 1 \leq j \leq \frac{N}{\tau}
\label{eq:coarse_grain}
\end{equation}
For scale one, the time series $\{y^{(1)}\}$ is simply the original time series. The length of each coarse-grained time series is equal to the length of the original time series divided by the scale factor $\tau$, as shown in Figure \ref{fig:multiscaling}.

\subsection{Refined composite multi-scale entropy (RCMSE)}
\label{subsec3}
\begin{figure}[H]
\centering 
\includegraphics[width=0.7\textwidth]{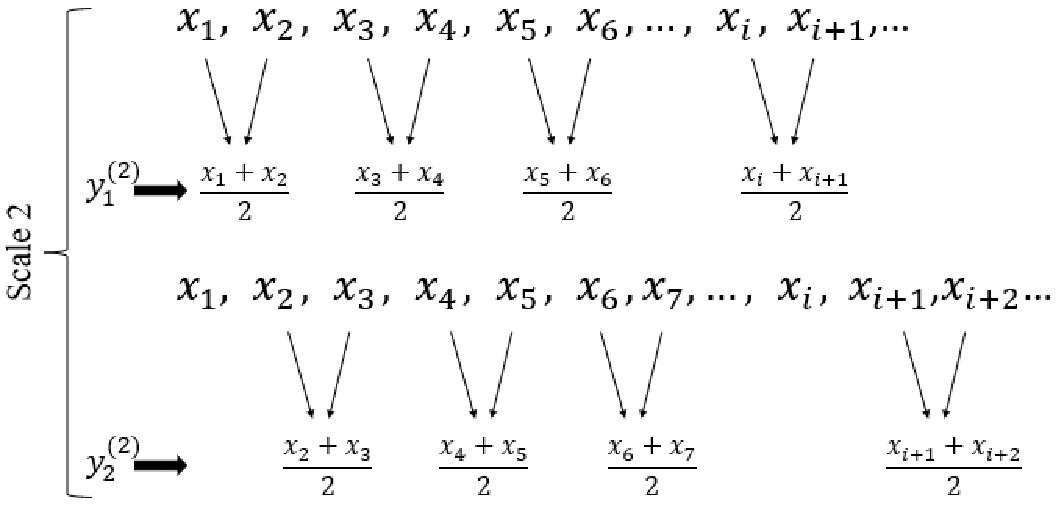}
\vspace{0.5cm}
\includegraphics[width=0.8\textwidth]{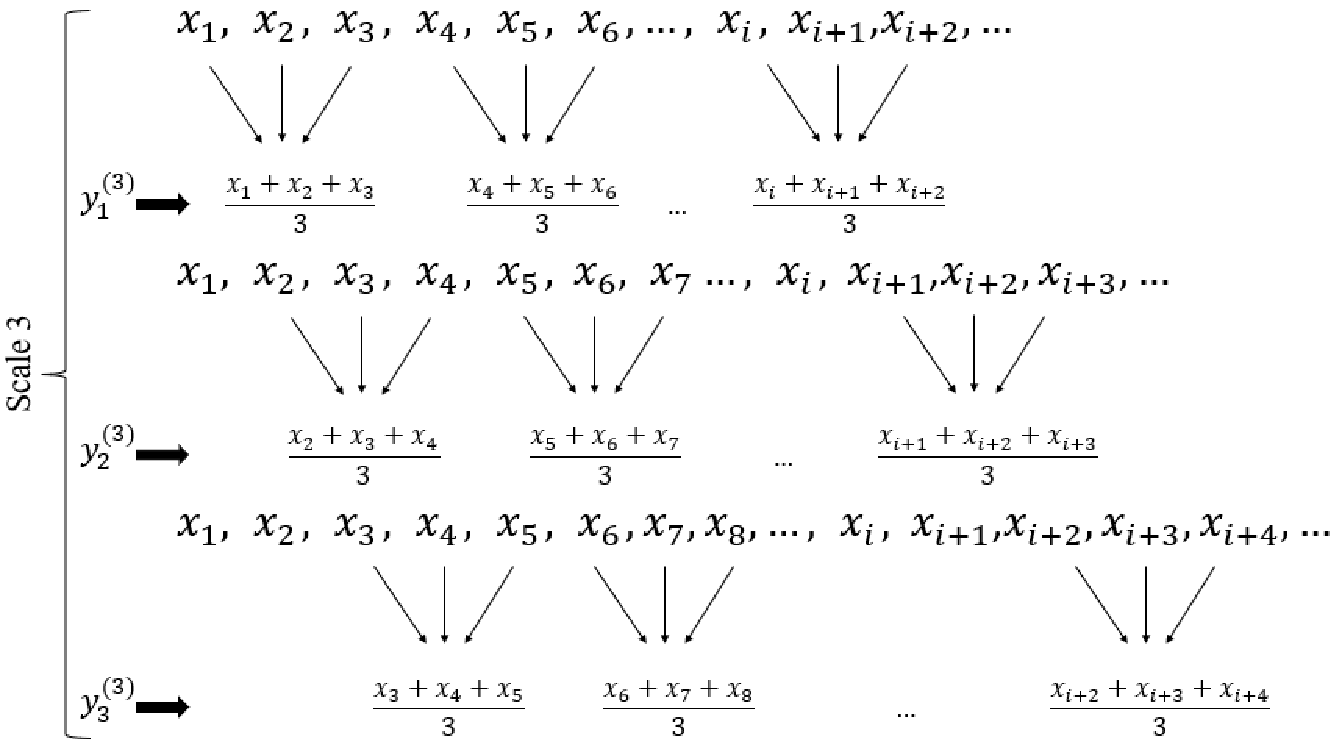}
\caption{Modified version of coarse graining procedure in RCMSE.}
\label{fig:scaling_images}
\end{figure}

A refined version of MSE was introduced in 2014 as RCMSE \cite{bib25}, which makes it more suitable for short time series. The new method depicted in Figure \ref{fig:scaling_images} divides the original time series into coarse-grained time series based on a specific scale factor, denoted as $\tau$. For a given time series $x = \{x_1, \dots, x_N\}$, the $k$-th coarse-grained time series $y_k^{(\tau)} = \{y_{(k,1)}^{(\tau)}, y_{(k,2)}^{(\tau)}, \dots, y_{(k,p)}^{(\tau)}\}$ is defined as follows:

\begin{equation}
y_{(k,j)}^{(\tau)} = \frac{1}{\tau} \sum_{i=(j-1)\tau+k}^{j\tau+k-1} x_i, \quad 1 \leq j \leq \frac{N}{\tau}, \quad 1 \leq k \leq \tau
\label{eq:new_coarse_grain}
\end{equation}

For scale one, as before, the time series is simply the original time series. At a scale factor of $\tau$, the number of matched vector pairs, $n_{(k,\tau)}^{m+1}$ and $n_{(k,\tau)}^m$, is calculated for all $\tau$ coarse-grained series. Let $\bar{n}_{(k,\tau)}^m$ ($\bar{n}_{(k,\tau)}^{m+1}$) represent the mean of $n_{(k,\tau)}^m$ ($n_{(k,\tau)}^{m+1}$) for $1 \leq k \leq \tau$. The RCMSE at a scale factor of $\tau$ is provided as Equation \eqref{eq:rcmse_equation}:

\begin{equation}
\text{RCMSE}(x,\tau,m,r) = -\ln\left(\frac{\bar{n}_{(k,\tau)}^{m+1}}{\bar{n}_{(k,\tau)}^m}\right) 
= -\ln\left(\frac{\sum_{k=1}^{\tau} n_{(k,\tau)}^{m+1}}{\sum_{k=1}^{\tau} n_{(k,\tau)}^m}\right)
\label{eq:rcmse_equation}
\end{equation}
Based on the above equation, the RCMSE value is undefined only when all $n_{(k,\tau)}^{m+1}$ or $n_{(k,\tau)}^m$ are zero.

\subsection{Multifractal Detrended Fluctuation Analysis (MF-DFA)}
\label{subsec4}
MF-DFA is a method used to analyze the multifractal characteristics of a time series. For a given time series $x=\{x_1, \dots, x_N\}$, this method can be described through the following steps \cite{bib21}:

Step 1) Compute the profile series $Y(i)$ based on the cumulative sum of the time series deviations from the mean:
\begin{equation}
Y(i) = \sum_{k=1}^i (x_k - \langle x \rangle), \quad i=1,\dots,N
\label{eq:cumulative_sum}
\end{equation}

Step 2) Segment the profile $Y(i)$ into $N_s = \text{int}(N/s)$ non-overlapping segments, each of length $s$. Because the length $N$ of the series is usually not an exact multiple of the chosen time scale $s$, a small portion at the end of the profile may be left over. To ensure this part is included, repeat the procedure starting from the opposite end. This results in a total of $2N_s$ segments.

Step 3) Determine the local trend for each of the $2N_s$ segments by fitting the series using least squares. Next, calculate the variance.

For $v = 1,2,\dots,N_s$:
\begin{equation}
F^2(s,v) = \frac{1}{s} \sum_{i=1}^s \left\{ Y[(v-1)s + i] - y_v(i) \right\}^2
\label{eq:cumulative_sum}
\end{equation}

For $v = N_s+1,\dots,2N_s$:
\begin{equation}
F^2(s,v) = \frac{1}{s} \sum_{i=1}^s \left\{ Y[N - (v-N_s)s + i] - y_v(i) \right\}^2
\label{eq:cumulative_sum}
\end{equation}
Here, $y_v(i)$ represents the fitting polynomial in segment $v$. The fitting procedure can employ linear, quadratic, cubic, or higher-order polynomials, commonly referred to as DFA1, DFA2, DFA3, and so on. This is denoted by $m$-order DFA, which means that each $m$-th order DFA removes a different trend in the time series.

Step 4) Compute the $q$-th order fluctuation function by averaging over all segments:
\begin{equation}
F_q(s) = \left\{ \frac{1}{2N_s} \sum_{v=1}^{2N_s} \left[ F^2(s,v) \right]^{q/2} \right\}^{1/q}
\label{eq:cumulative_sum}
\end{equation}

In the above equation, $q$ can be any real value except 0. Positive values of $q$ emphasize segments with larger fluctuations, making the fluctuation function $F_q(s)$ more sensitive to large-scale deviations from the local trend, while negative values of $q$ highlight segments with smaller fluctuations, making $F_q(s)$ more sensitive to smaller deviations from the local trend. When $q = 2$, the result is the standard DFA.

Step 5) Analyze the scaling behavior of the fluctuation functions by examining the log-log plots of $F_q(s)$ against $s$ for each $q$ value. If the series $x(i)$ exhibit long-range power-law correlations, $F_q(s)$ will increase as a power-law for large values of $s$:

\begin{equation}
F_q(s) \sim s^{h(q)}
\label{eq:cumulative_sum}
\end{equation}

where $h(q)$ is called the generalized Hurst exponent. The multifractal characteristics of time series are also examined through multifractal spectrum analysis, which relies on the analytical relationship between the generalized Hurst exponents $h(q)$ and the Rényi exponent $\tau(q)$. The exponent $\tau(q)$, also known as the mass exponent or multifractal scaling exponent, is a function that characterizes how the moments of a signal or time series scale with its observation scale $s$. A time series is characterized as multifractal if $\tau(q)$ is a nonlinear function of $q$. The relation between $\tau(q)$ and $h(q)$ is described with the equation \eqref{eq:tau_equation}.

\begin{equation}
\tau(q) = q h(q) - 1
\label{eq:tau_equation}
\end{equation}

The exponent $\tau(q)$ captures how different statistical moments (such as averages and variances) of fluctuations in the series scale with the segment size $s$. To derive a singularity spectrum $f(\alpha)$
from MF-DFA based on standard multifractal analysis \cite{bib27}, it is necessary to determine the local singularity strength, or Hölder exponent $\alpha$, from the generalized Hurst exponent $h(q)$. The local singularity strength $\alpha$ and singularity spectrum $f(\alpha)$
, associated with the power $q$, can be calculated as:

\begin{equation}
\alpha(q) = \frac{d \tau(q)}{dq}, \quad f(\alpha) = q \alpha - \tau(q)
\label{eq:alpha_equation}
\end{equation}

The exponent $\alpha$ measures how rapidly fluctuations diminish as the length of these segments decreases. Since $\alpha$ is positive, a larger $\alpha$ value means a quicker decay of fluctuations as $s \to 0$, indicating a smoother time series in that region. Consequently, smaller values of $\alpha$ correspond to rougher segments in the time series. The singularity spectrum $f(\alpha)$ is the Legendre transformation of $\tau(q)$. The function $f(\alpha)$ is a convex function of $\alpha$ with a peak value of 1 for any one-dimensional time series. This peak, occurring at $q = 0$, represents the capacity or topological dimension of the time series, which is always 1 for one-dimensional cases. The width of the singularity spectrum $f(\alpha)$
, defined as $\Delta \alpha = \alpha_{\text{max}} - \alpha_{\text{min}}$, reflects the range of Hölder exponents needed to characterize the time series. A broader range of $\alpha$ values corresponds to greater complexity within the time series. The parameters $\alpha$ and $f(\alpha)$ can be expressed directly in terms of $h(q)$ as follows:  

\begin{equation}
\alpha = h(q) + qh'(q), \quad f(\alpha) = q[\alpha - h(q)] + 1
\end{equation}

\section{Data Description}\label{sec3}

\begin{figure}[H]
\centering
\includegraphics[width=1\textwidth]{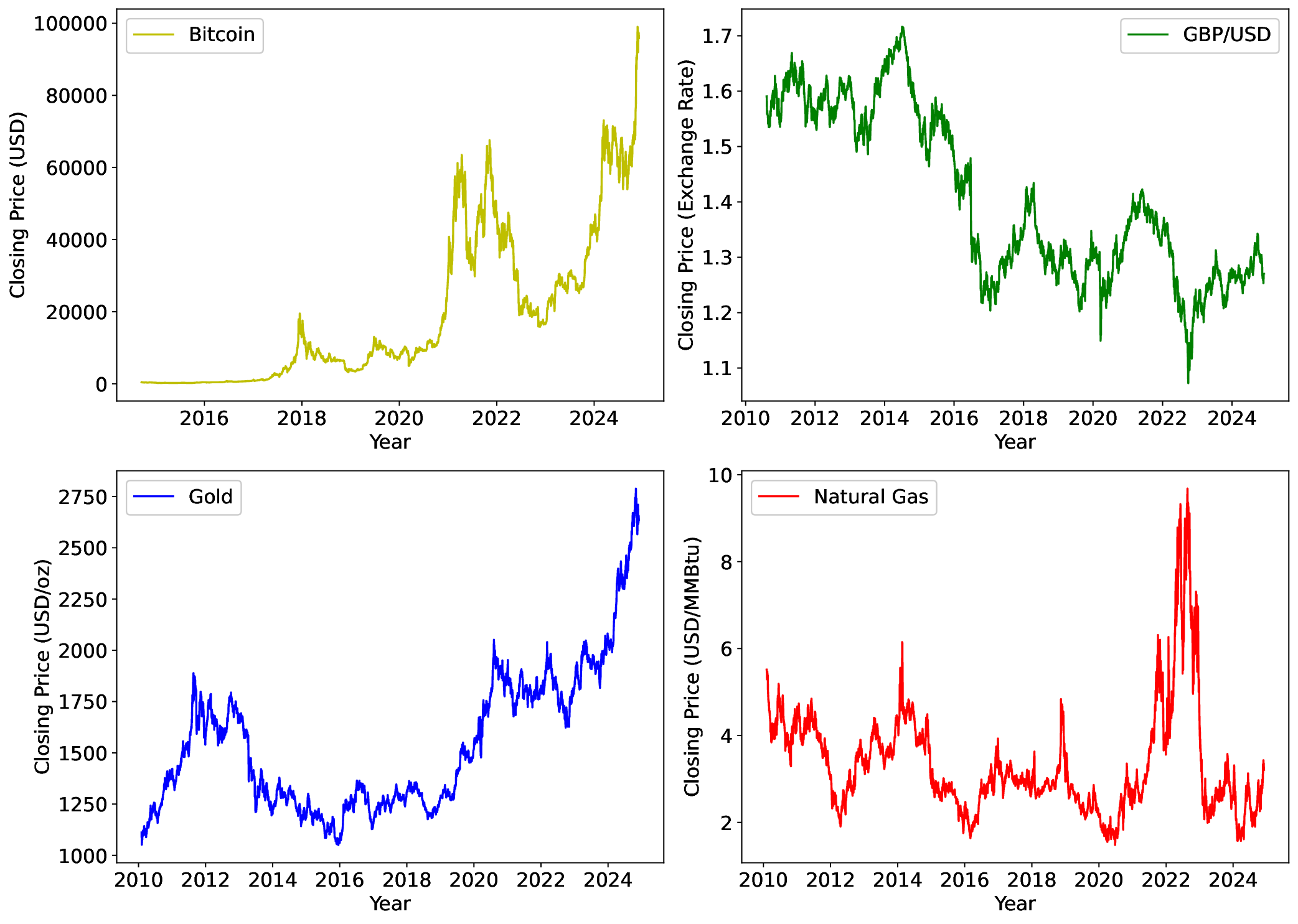}
\caption{Historical daily closing price data for Bitcoin, GBP/USD, gold, and natural gas, covering up to December 2, 2024. Each time series contains 3,730 data points.}
\label{fig:time_series}
\end{figure}

This study uses the daily price data of four key financial assets: Bitcoin, GBP/USD, gold, and natural gas. The reason for choosing these assets is that they cover a spectrum of asset types—cryptocurrencies, forex, commodities, and precious metals—each offering distinct volatility profiles, market drivers, and investor behaviors. This diversity helps us examine how complexity varies across assets with different economic roles and investor perceptions. The data were sourced from the \href{https://finance.yahoo.com/}{Yahoo Finance website}, providing daily price reports.The datasets cover different time spans based on the availability of data for each asset, extending up to 2024-12-02. All data have the same length, with 3,730 data points. Fig.~\ref{fig:time_series} presents the price time series of each asset, capturing long-term trends and fluctuations over their respective periods.

In this study, the logarithmic returns of the data were used instead of the original price data, as they more accurately capture market fluctuations, which are indicative of risk. Logarithmic returns were calculated on the basis of closing prices according to the following equation:

\begin{equation}
    \text{Logarithmic Return} = \ln(P_t) - \ln(P_{t-1})
= \ln\left(\frac{P_t}{P_{t-1}}\right)
    \label{eq:log_return}
\end{equation}

\begin{figure}[H]
    \centering
    \includegraphics[width=1\textwidth]{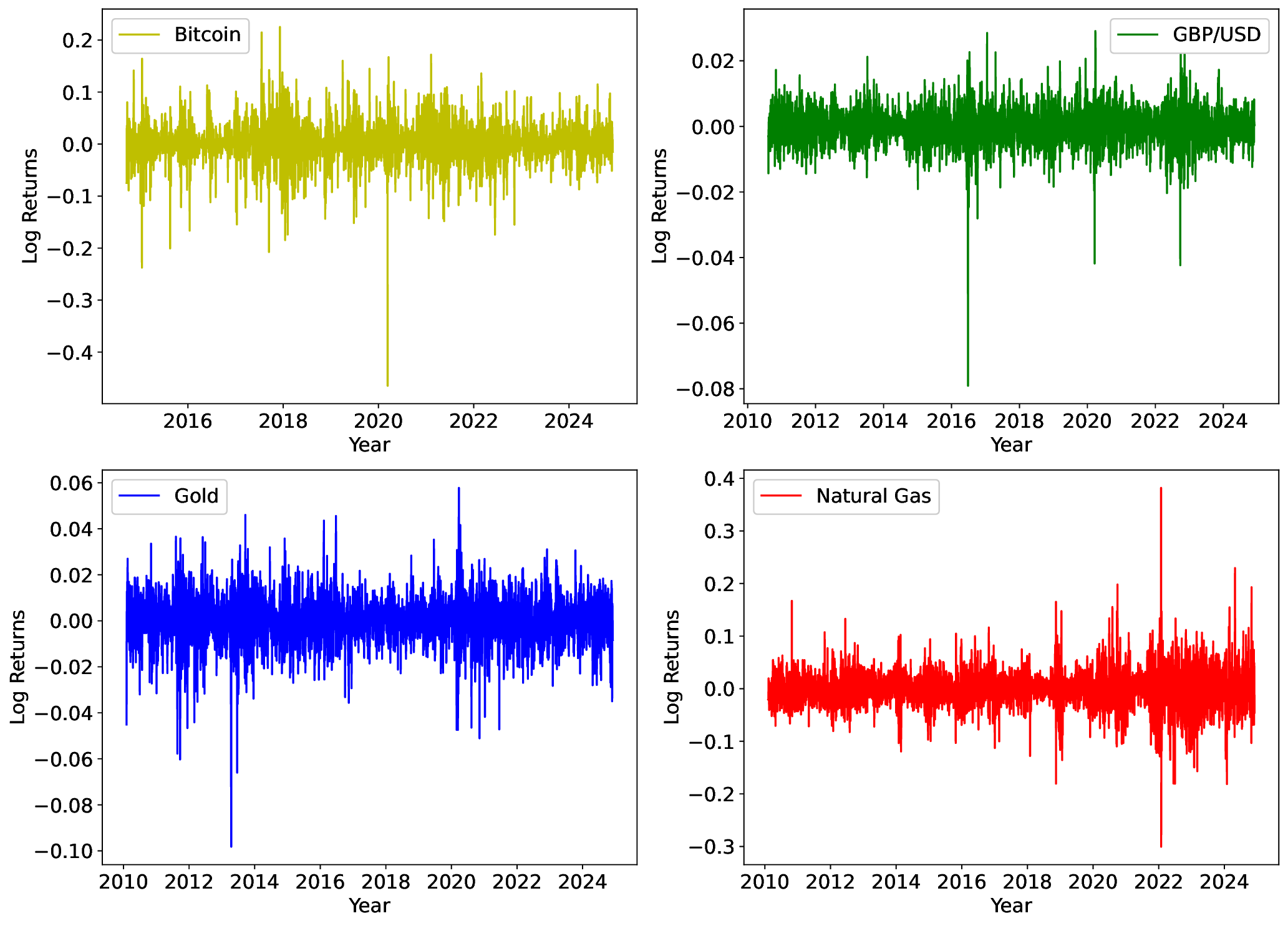}
    \caption{Logarithmic returns of Bitcoin, GBP/USD, gold, and natural gas calculated based on their closing prices. The plot illustrates the daily fluctuations of these assets, highlighting their market volatility and risk dynamics.}
    \label{fig:log_returns}
\end{figure}

Where $P_t$ is the price at time $t$ and $P_{t-1}$ is the price at the previous time step. Fig.~\ref{fig:log_returns} illustrates the logarithmic return time series for each asset over the specified periods based on the closing price. The visual representation highlights significant volatility and fluctuation patterns across different time periods, which are characteristic of the behavior observed in financial markets. In particular, the logarithmic return series reveal episodes of extreme price movements, reflecting market dynamics influenced by a range of economic factors.

\section{Results}\label{sec4}
\subsection{Data Distribution}
\begin{figure}[H]
    \centering
    \includegraphics[width=1\textwidth]{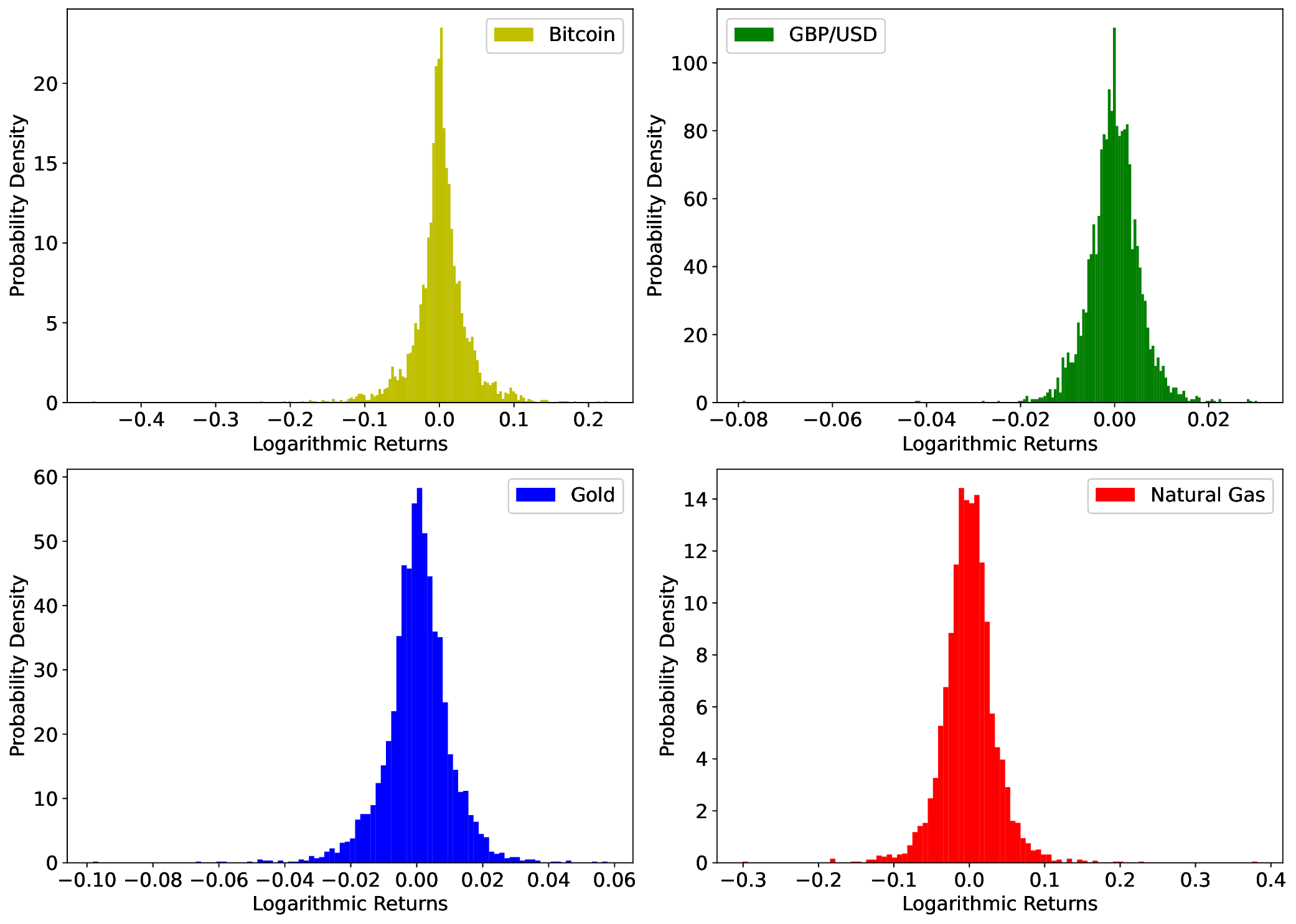}
    \caption{Probability density function of the logarithmic returns of Bitcoin, GBP/USD, gold, and natural gas. The wider distributions of Bitcoin and natural gas indicate higher volatility in these markets compared to GBP/USD and gold.}
    \label{fig:pdf_plot}
\end{figure}

To enhance the understanding of data fluctuations and distribution, a probability density function (PDF) helps deepen comprehension of the data. The PDFs of logarithmic returns are presented in Fig.~\ref{fig:pdf_plot}.
The wider distribution of Bitcoin and natural gas indicates higher volatility in these assets. GBP/USD and gold exhibit narrower distributions, reflecting smaller fluctuations in those assets.

The deviation of probability density functions from a normal distribution can be evaluated by the skewness and excess Kurtosis defined by:

\begin{equation}
\text{Skewness} = \frac{\mu_3}{\sigma^3},
\label{eq:skewness}
\end{equation}
 
and 

\begin{equation}
\text{ex-Kurtosis} = \frac{\mu_4}{\sigma^4} - 3.
\label{eq:excess_kurtosis}
\end{equation}
In the above equations, $\sigma$ is the standard deviation of the distribution. 
$\mu_3$ and $\mu_4$ are the third and fourth moments, respectively, and given by:

\begin{equation}
\mu_3 = \frac{1}{n} \sum_{i=1}^n (x_i - \bar{x})^3
\label{eq:third_moment}
\end{equation}

\begin{equation}
\mu_4 = \frac{1}{n} \sum_{i=1}^{n} (x_i - \bar{x})^4,
\label{eq:fourth_central_moment}
\end{equation}
where $\bar{x}$ is the mean of data and $n$ is the data length \cite{bib28}.

Table~\ref{kurtosis and skewness} shows the skewness, ex-kurtosis, and standard deviation of the logarithmic returns distribution. The non-zero skewness indicates the non-symmetry of the probability density function.
Positive skewness of gas suggests a tendency toward more frequent positive returns, while Bitcoin, GBP/USD, and gold exhibit negative skewness values, indicating that downward price movements are more common.

The high kurtosis across all assets reflects the 
non-normal nature of financial returns, with frequent extreme values indicating substantial risks. Among the assets, GBP/USD shows the highest kurtosis value, indicating heavy tails in its distribution. This suggests the presence of frequent and significant deviations from the mean, possibly due to sudden reactions in the currency market to political events and economic news.

\begin{table}[h]
\caption{Skewness, ex-Kurtosis, and standard deviation values for the log-return time series of Bitcoin, GBP/USD, gold, and natural gas.}\label{kurtosis and skewness}%
\begin{tabular}{@{}lllll@{}}
\toprule
 & Bitcoin  & GBP/USD & gold & gas\\
\midrule
Skewness    & -0.72   & -0.88  & -0.59 & 0.42  \\
ex-Kurtosis    & 11.36   & 14.26  & 5.91 & 6.78\\
Standard Deviation  & 0.036   & 0.005  & 0.010 & 0.035\\
\botrule
\end{tabular}
\end{table}

%Kurtosis and Skewness are computed according to Equations \eqref{eq:excess_kurtosis} and \eqref{eq:skewness}, respectively\cite{bai2005tests}.

\subsection{RCMSE analysis} 
\begin{figure}[H] 
    \centering
    \includegraphics[width=0.7\textwidth]{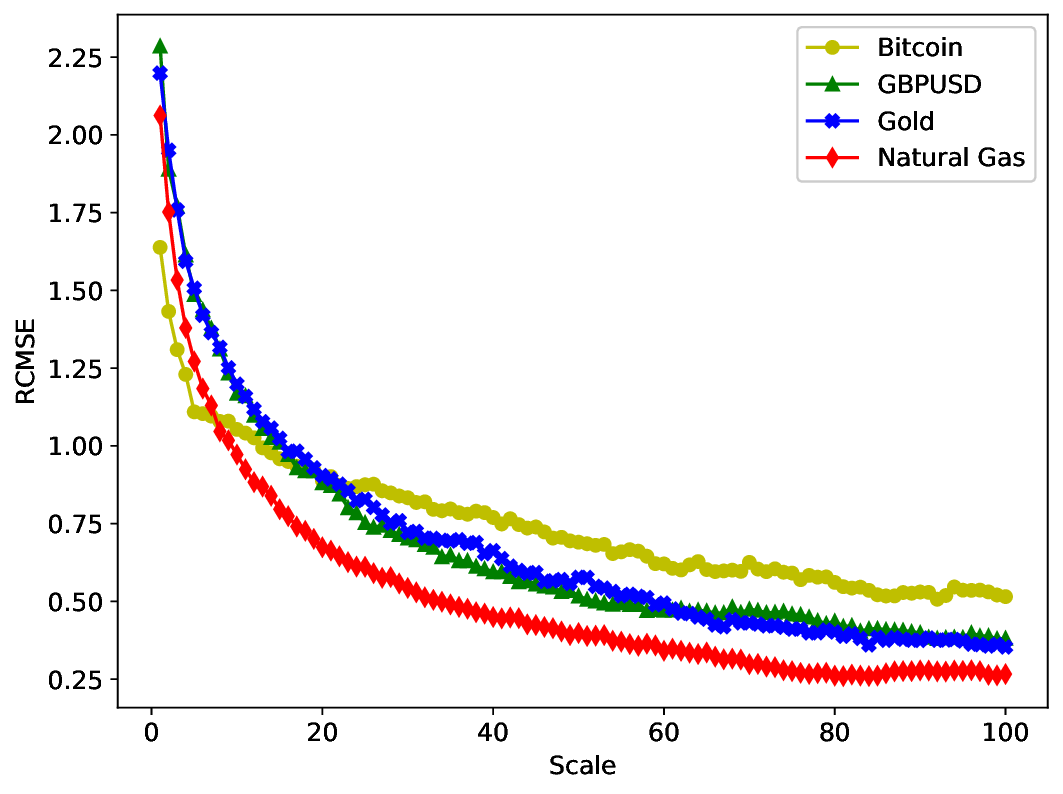} % Set width and height
    \caption{RCMSE across scales for the log returns of Bitcoin, GBP/USD, gold, and natural gas. Bitcoin exhibits the lowest entropy values at the smaller scales and the highest entropy values at the larger scales.}
    \label{fig:RCMSE}
\end{figure}

As mentioned, RCMSE—a refined version of MSE—is particularly suitable for short time series, especially at larger scales. Common MSE often results in undefined values for short time series at large scales, but this challenge is addressed by RCMSE. As shown in Figure \ref{fig:scaling_images}, at a scale factor of $\tau$, multiple coarse-grained series are generated, reducing the probability of undefined values in Equation~\ref{eq:rcmse_equation}. The RCMSE analysis of the logarithmic returns for Bitcoin, GBP/USD, gold, and natural gas over 100 coarse-grained scales is shown in Figure~\ref{fig:RCMSE}. For this analysis, the embedding dimension m was set at 3, and the tolerance r was defined as 0.15 times the standard deviation of the time series. It can be observed that entropy decreases as the scale factor increases across all scales. However, Bitcoin exhibits the lowest entropy on the first scales and the highest entropy on larger scales compared to other assets. This suggests that Bitcoin is more predictable on low scales, whereas GBP/USD has the highest entropy and unpredictability at low scales. 

A quantitative measure of complexity is obtained by summing the entropy values across all scales. Table~\ref{tab:complexity} presents the complexity values for all assets. The findings indicate that Bitcoin generally exhibits the highest complexity, despite having the lowest entropy values at smaller scales. In contrast, gas shows the lowest overall complexity among the analyzed assets. It is worth mentioning that a higher complexity value does not necessarily imply lower predictability at a specific time scale; rather, it provides insight into average predictability across multiple time scales. Higher complexity reflects greater irregularity and volatility across all time scales, not just one.

\begin{table}[h]
\caption{The complexity values across 100 time scales, as well as the entropy at the first scale, were computed for Bitcoin, GBP/USD, Gold, and Gas.}\label{tab:complexity}%
\begin{tabular}{@{}lllll@{}}
\toprule
 & Bitcoin  & GBP/USD & gold & gas\\
\midrule
Complexity    & 74.66\textsuperscript{*}   & 67.24  & 67.88 & 51.48  \\
\botrule
\end{tabular}
\end{table}

\begin{figure}[H] 
    \centering
    \includegraphics[width=0.7\textwidth]{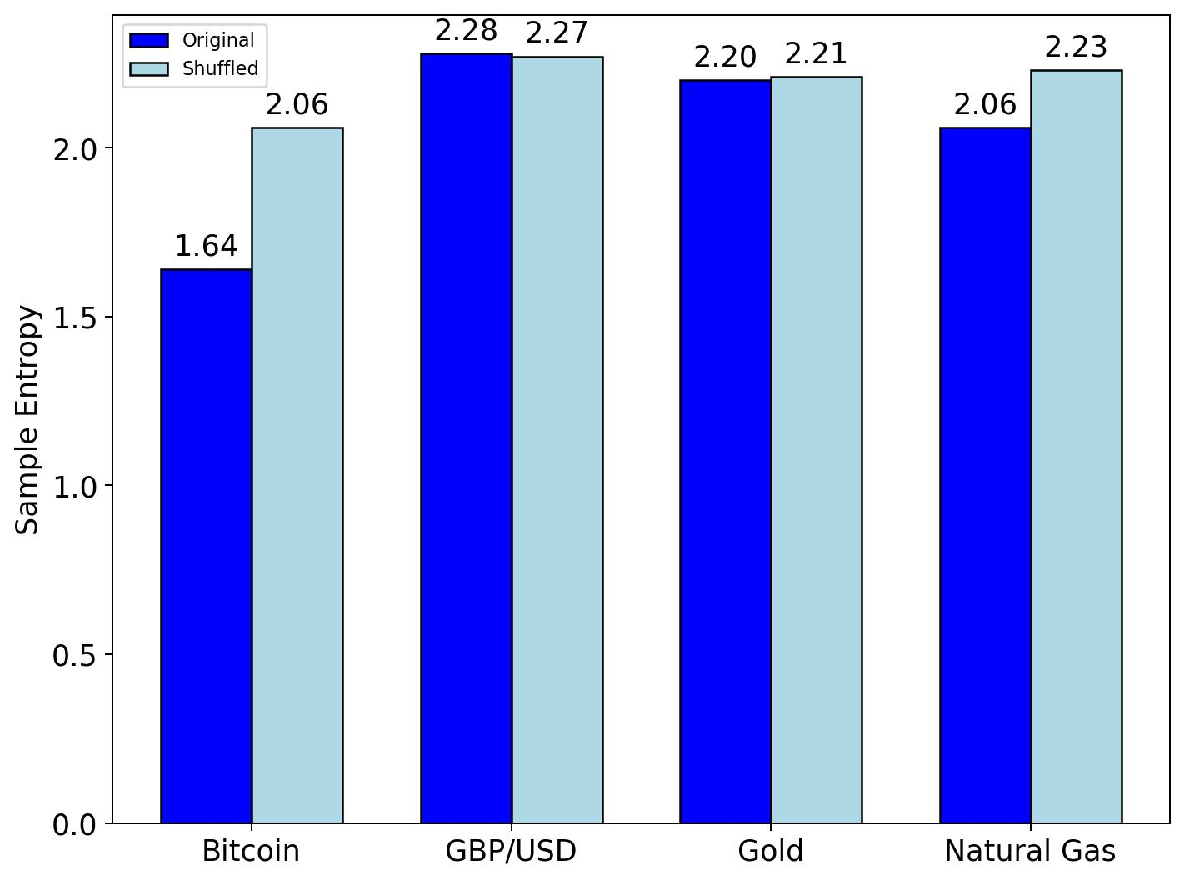} % Set width and height
    \caption{The sample entropy of Bitcoin, GBP/USD, Gold, and Gas was computed before and after shuffling. Bitcoin shows a notable change in entropy after shuffling, indicating a significant alteration in its temporal structure.}
    \label{fig:Barchart_Entropy}
\end{figure}

We computed the sample entropy of both the original and shuffled time series to better understand and compare the regularity of the data. Entropy values for all assets, before and after shuffling, are presented as a bar chart in Figure~\ref{fig:Barchart_Entropy}.
Each series was shuffled 100 times using a computer-based randomization process to eliminate temporal dependencies. The increase in the entropy value of Bitcoin compared to the unshuffled series indicates the presence of non-linear correlations and more regularity in its original data. In contrast, no significant changes were observed in the entropy values of the other assets before and after shuffling, indicating their inherently random characteristics.

\subsection{MF-DFA analysis}
\begin{figure}[H] 
    \centering
    \includegraphics[width=0.7\textwidth]{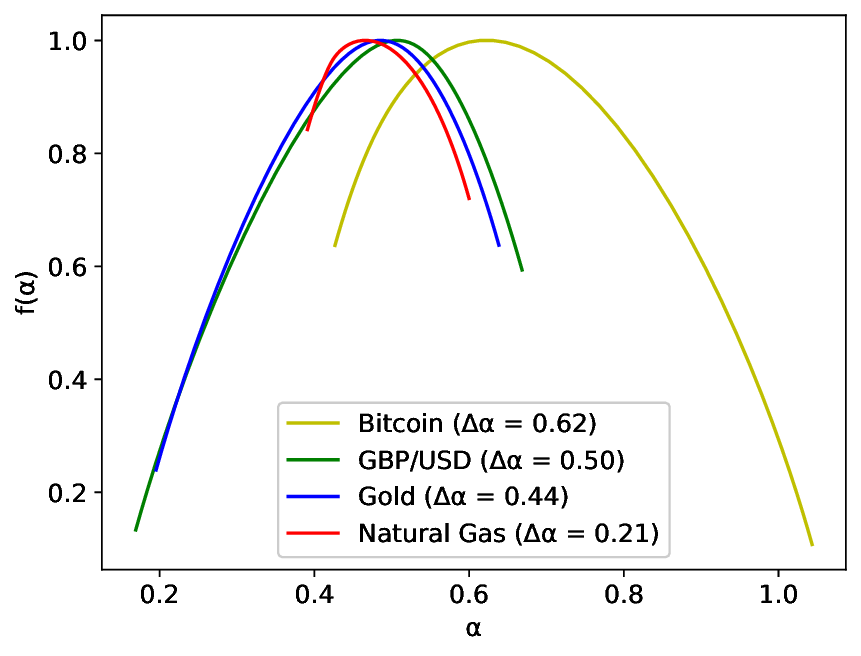} % Set width and height
    \caption{Singularity spectrum of Bitcoin, GBP/USD, gold, and natural gas. The wider spectrum of Bitcoin indicates its greater complexity, while natural gas exhibits the lowest complexity.}
    \label{fig:MFDFA}
\end{figure}

MF-DFA analysis shows the fractal behavior of a time series. Unlike traditional monofractal analysis, which only considers a single-scale behavior, multifractal analysis reflects the fractal nature of a series over different scales. Financial time series, such as those of Bitcoin, GBP/USD, gold, and natural gas, often exhibit complex dynamics due to their multifractal behavior. To quantify this complexity, we applied MF-DFA, which allows us to assess how fluctuations at different time scales contribute to the overall structure of the market. One of the most important analyses in MF-DFA is the multifractal spectrum, or singularity spectrum. A wider singularity spectrum indicates greater complexity and a higher degree of fractality in the corresponding data. Note that complexity in this context is not the same as entropy. Rather, complex behavior in terms of multifractality reflects the presence of fluctuations occurring at multiple frequencies across different time scales. The multifractal spectrum of assets is shown in Figure~\ref{fig:MFDFA}. The results show a wider spectrum for Bitcoin compared to the other assets, indicating greater complexity and richer multifractal nature. In contrast, the smallest spectrum is associated with natural gas. Table~\ref{tab:spectrum width}, presents the width of the multifractal spectrum of all assets.

\begin{table}[h]
\caption{The multifractal spectrum widths of Bitcoin, GBP/USD, and gold are shown, with Bitcoin exhibiting a notably different spectrum compared to the others.}\label{tab:spectrum width}%
\begin{tabular}{@{}lllll@{}}
\toprule
 & Bitcoin  & GBP/USD & Gold & Gas\\
\midrule
Spectrum Width    & 0.62\textsuperscript{*}   & 0.50  & 0.44 & 0.21  \\
\midrule
$\alpha_{\text{max}}$ & 1.04 & 0.67 & 0.64 & 0.60 \\
\midrule
$\alpha_{\text{min}}$ & 0.43 & 0.17 & 0.19 & 0.39 \\
\botrule
\end{tabular}
\end{table}

Small values of $\alpha$ correspond to large fluctuations, higher volatility, and abrupt dynamics, while large $\alpha$ values indicate small fluctuations and smoother behavior with gradual variations. As presented in Table~\ref{tab:spectrum width}, Bitcoin's spectrum covers larger $\alpha$ values compared to the other assets. This indicates a more regular and smoother behavior in Bitcoin's data, which is also confirmed by its lowest entropy.

\begin{figure}[H] 
    \centering
    \includegraphics[width=0.7\textwidth]{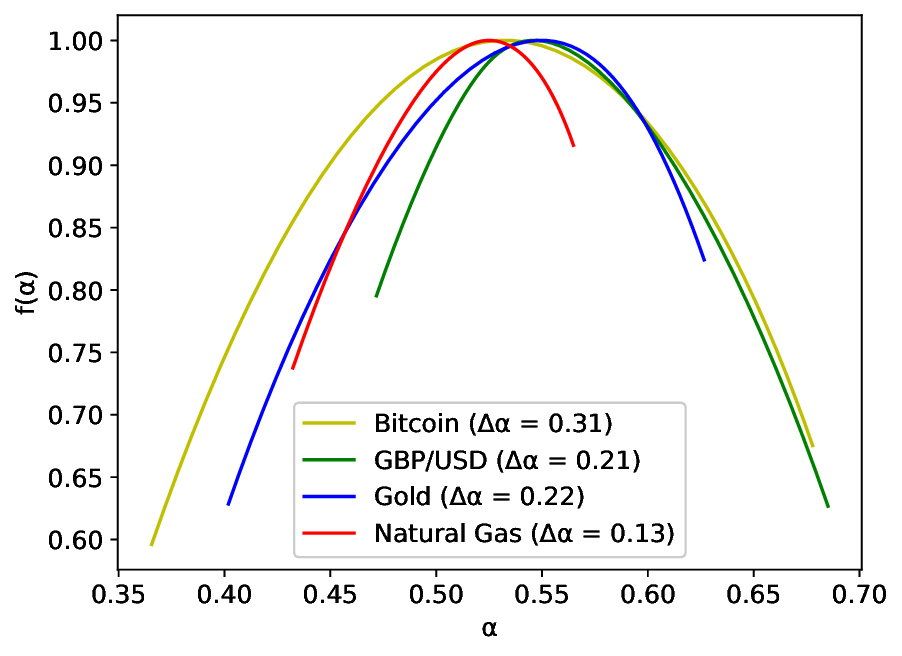} % Set width and height
    \caption{Multifractal spectrum of the shuffled data for Bitcoin, GBP/USD, Gold, and Natural Gas. A notable reduction in the width of the spectrum for all assets is observed compared to the original data.}
    \label{fig:MFDFA_shuffled}
\end{figure}

In another analysis, we calculated the multifractal spectrum of the shuffled series for all assets to eliminate potential long-term dependencies and regularities in the data, in order to better understand the degree of multifractality. Figure~\ref{fig:MFDFA_shuffled} presents the multifractal spectra of the shuffled data for all assets. A notable reduction in the spectra was observed across all assets, which may indicate a rich degree of multifractality in the original, unshuffled data.

\begin{figure}[H] 
    \centering
    \includegraphics[width=0.7\textwidth]{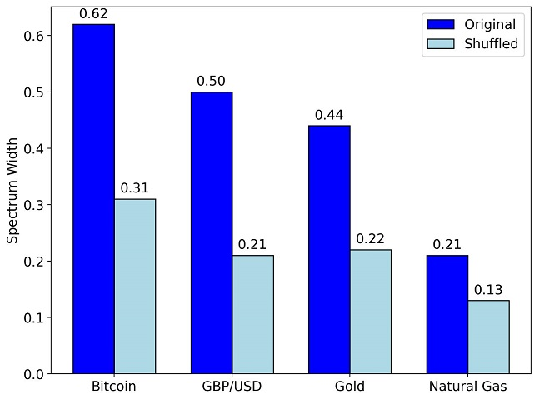} % Set width and height
    \caption{Singularity spectrum width of both shuffled and unshuffled data for Bitcoin, GBP/USD, Gold, and Gas. A notable reduction in the spectrum width of the shuffled data can be observed compared to the unshuffled data.}
    \label{fig:Barchart_singualrity_spectrum}
\end{figure}

Figure~\ref{fig:Barchart_singualrity_spectrum} presents a bar chart comparing the singularity width of both shuffled and unshuffled data for all assets.

While the Hurst exponent values obtained from DFA, being close to 0.5 for the assets, indicate a lack of linear autocorrelation in the data, the notable reduction in the singularity spectrum of the shuffled data compared to the original data may indicate the presence of non-linear autocorrelations, which are not captured by the Hurst exponent.

\begin{figure}[H] 
    \centering
    \includegraphics[width=1\textwidth]{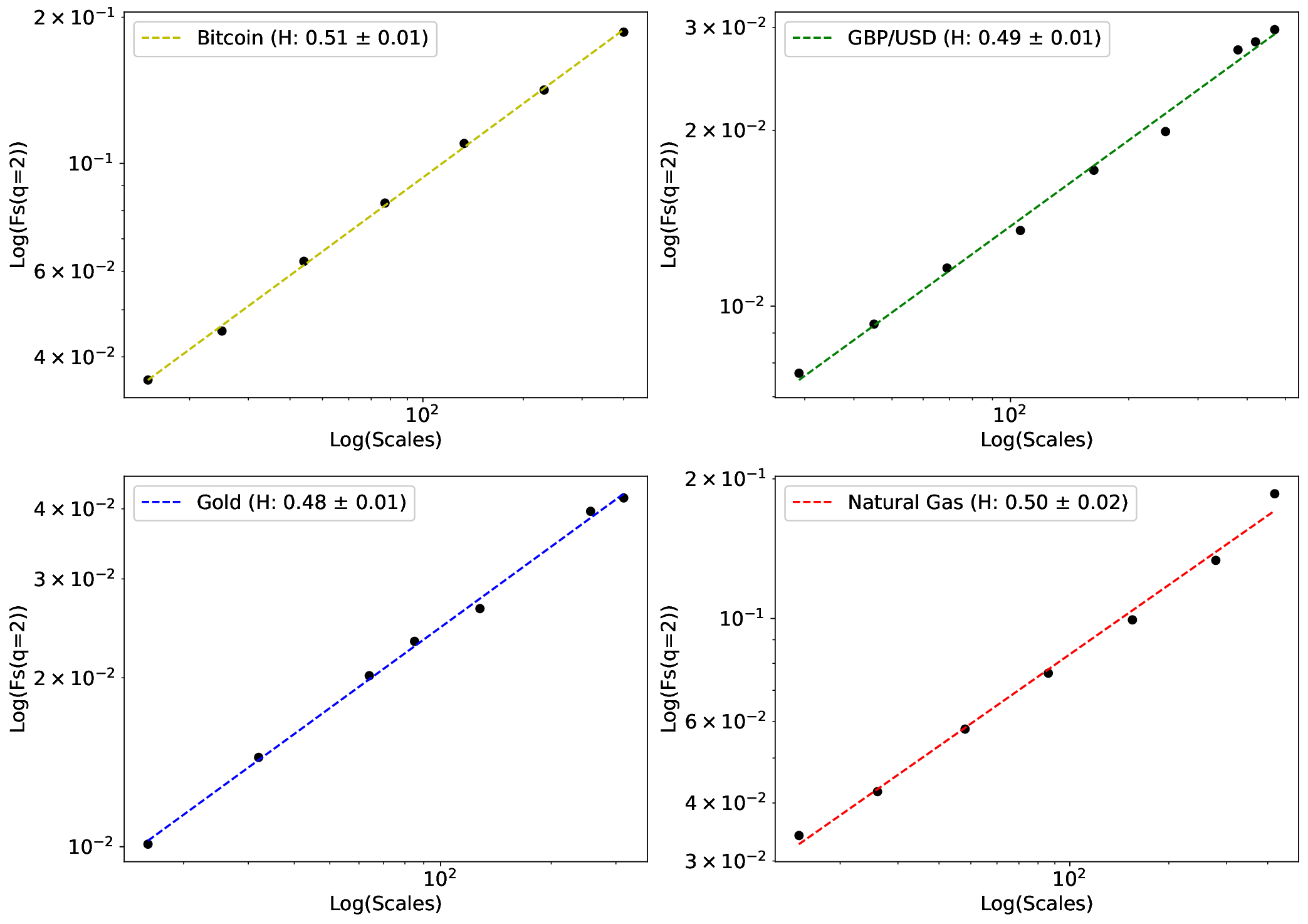} % Set width and height
    \caption{Log-log plot of the fluctuation function over scale for $q=2$ for Bitcoin, GBP/USD, gold, and natural gas. The slope of the fitted line represents the Hurst exponent.}
    \label{fig:Hurst}
\end{figure}

Figure~\ref{fig:Hurst} shows the log-log plot of the fluctuation function over scales. The slope of the fitted line in this plot for $q = 2$ corresponds to the Hurst exponent. The Hurst exponents of the assets can also be seen in Table~\ref{tab:Hurst}.

\begin{table}[h]
\caption{The Hurst exponents of Bitcoin, GBP/USD, Gold, and Gas are obtained as follows. Values close to 0.5 indicate the absence of significant linear autocorrelation in their data.}\label{tab:Hurst}%
\begin{tabular}{@{}lllll@{}}
\toprule
 & Bitcoin  & GBP/USD & Gold & Gas\\
\midrule
Hurst Exponent    & $0.51 \pm 0.1$ & $0.49 \pm 0.1$  & $0.48 \pm 0.1$ & $0.50 \pm 0.2$  \\
\botrule
\end{tabular}
\end{table}

By combining both entropy and multifractal analysis, the distinct behavior of Bitcoin compared to the other assets becomes evident. This distinction is reflected in its lower entropy value and higher degree of multifractality. While the Hurst exponent indicates no significant linear dependencies in Bitcoin's data, the presence of non-linear autocorrelations may still exist.

\section*{Conclusion}
In this research, we employed Multifractal Detrended Fluctuation Analysis (MF-DFA) and Refined Composite Multiscale Sample Entropy (RCMSE) to investigate the predictability and fractal behavior of financial time series. While the multiscale entropy analysis reveals the degree of predictability across different time scales, the multifractal analysis captures the fluctuating nature of the time series and highlights the presence of nonlinear correlations in the data. The overall findings of this research can be summarized as follows:

Both the Bitcoin and Gas log-return time series exhibit higher volatility compared to GBP/USD and Gold, as reflected in their probability density function plots. However, despite their similar volatility levels, they display distinct behaviors over short and long time scales.

Bitcoin exhibits the lowest entropy at smaller scales but shows higher entropy at larger scales compared to the other assets. It also exhibits the highest overall complexity, calculated as the sum of entropy values across all scales. These findings suggest that Bitcoin exhibits the most regular and predictable behavior at short time scales, but greater instability at larger time scales. Furthermore, Bitcoin exhibits the widest singularity spectrum, indicating a broad range of singularity exponents across different time scales.

All assets demonstrate Hurst exponents close to 0.5, indicating a lack of linear dependencies in their data. However, the larger increase in the entropy of shuffled Bitcoin log-return time series with respect to its original one, along with the reduction of its singularity spectrum widths upon shuffling, reflects the presence of nonlinear correlations in this asset.

This research demonstrates that financial time series can exhibit mixed characteristics over both short and long terms, as observed in Bitcoin's behavior, alongside strong multifractal properties. These findings reflect the deep and complex dynamics of financial systems, which may arise from a combination of factors such as market forces, individual decision-making, and political events.

\section*{Author Contributions}

\textbf{Oday Masoudi}: Writing – original draft, Programming, Visualization, Validation, Methodology, Investigation, Data curation.

\textbf{Farhad Shahbazi}: Writing – original draft, Results interpretation, Validation, Supervision, Conceptualization.

\textbf{Mohammad Sharifi}: Methodology, Investigation, Programming.

\bmhead{Acknowledgements}

The authors would like to thank Hamed Farahani for his valuable guidance on the writing process and for sharing helpful reference materials during the preparation of this paper.
%%===========================================================================================%%
%% If you are submitting to one of the Nature Portfolio journals, using the eJP submission   %%
%% system, please include the references within the manuscript file itself. You may do this  %%
%% by copying the reference list from your .bbl file, paste it into the main manuscript .tex %%
%% file, and delete the associated \verb+\bibliography+ commands.                            %%
%%===========================================================================================%%

\bibliography{sn-bibliography}% common bib file
%% if required, the content of .bbl file can be included here once bbl is generated
%%\input sn-article.bbl

\end{document}